\RequirePackage{amsmath}
\documentclass[12pt]{iopart}

\usepackage{array,url}
\usepackage{arydshln,amsthm,amssymb}
\usepackage{graphicx}
\usepackage{bm}

\newcommand{\ket}[1]{\left|{#1}\right\rangle}

\newcommand{\ip}[2]{\left\langle{#1}\right|\left.\!{#2}\right\rangle}

\begin{document}

\title[]{Scrambling for precision: optimizing multiparameter qubit estimation in the face of sloppiness and incompatibility}
\author{Jiayu He$^1$ and Matteo G. A. Paris$^2$}
\address{$^1$ QTF Centre of Excellence, Department of Physics, University of Helsinki, FI-00014 Helsinki, Finland}
\address{$^2$ Dipartimento di Fisica {\em Aldo Pontremoli}, Universit\`a degli Studi di Milano, I-20133 Milano, Italy}
\ead{jiayu.he@helsinki.fi, matteo.paris@fisica.unimi.it}
\date{\today}
\begin{abstract}
Multiparameter quantum estimation theory plays a crucial role in advancing quantum metrology. Recent studies focused on fundamental challenges such as enhancing precision in the presence of incompatibility or sloppiness, yet the relationship between these features remains poorly understood. In this work, we explore the connection between sloppiness and incompatibility by introducing an adjustable scrambling operation for parameter encoding. Using a minimal yet versatile two-parameter qubit model, we examine the trade-off between sloppiness and incompatibility and discuss: (1) how information scrambling can improve estimation, and (2) how the correlations between the parameters and the incompatibility between the symmetric logarithmic derivatives impose constraints on the ultimate quantum limits to precision. Through analytical optimization, we identify strategies to mitigate these constraints and enhance estimation efficiency. We also compare the performance of joint parameter estimation to strategies involving successive separate estimation steps, demonstrating that the ultimate precision can be achieved when sloppiness is minimized. Our results provide a unified perspective on the trade-offs inherent to multiparameter qubit statistical models, offering practical insights for optimizing experimental designs. 
\end{abstract}
\section{Introduction}
Multiparameter quantum estimation \cite{demkowicz2020multi,albarelli2020perspective,liu2020quantum,razavian2020quantumness,carollo2019,PhysRevResearch.4.043057,sidhu20} is an active area of research for its fundamental interest \cite{Bonalda2019,Danelli2025} and  due to its wide range of applications in quantum metrology \cite{tsang2011,baumgratz2016,Ansari2021}, quantum imaging \cite{Genovese2016,tsang2019,ang2017,vrehavcek2017}, and other fields\cite{vidrighin2014,crowley2014, pezze2017,gessner2018,proctor2018,gefen2019,chen2019,PhysRevLett.134.030802,belliardo2024}.
Additionally, the recent finding that simultaneous estimation of many parameters can yield a better precision limit than estimating each parameter individually has accelerated the development of this field \cite{Humphreys2013,yue2014,liu2016,gagatsos2016}. One of the most important targets in quantum multiparameter metrology is improving parameter estimation precision. The quantum Fisher information matrix (QFIM) and the accompanying quantum Cram$\acute{e}$r-Rao bound (QCRB) are relevant tools in this endeavor \cite{holevo2011,helstrom1969,caves1,caves2,braunstein1994,paris2009quantum}.

Multiparameter quantum estimation, similarly to the single-parameter case, involves three steps: probe state preparation, parameter encoding via system-probe interaction, and measurement-based information extraction. However, because of the correlation between parameters, designing optimal estimation strategies becomes more challenging. During encoding, a key challenge arises from {\em sloppiness}, which is a phenomenon where redundant or poorly encoded parameters reduce the efficiency of information extraction. Sloppiness \cite{brown2003statistical,brown2004statistical,PhysRevLett.97.150601,machta2013parameter,PRXQuantum.2.020308,Goldberg21,Yang23,Frigerio2024,PhysRevA.111.012414}
occurs when parameters are not independently encoded into the quantum probe state, leading to correlations that obscure individual parameter estimation, which acts as an intrinsic noise source. In contrast, {\em stiffness} refers to the desirable scenario where parameters are encoded in a way that minimizes redundancy, allowing for efficient and independent estimation of each parameter, thereby enhancing the overall precision of the estimation process.

Meanwhile, the measurement stage introduces a fundamental trade-off: optimizing precision for one parameter often compromises others due to the incompatibility of non-commuting observables. For instance, when the symmetric logarithmic derivatives (SLDs) corresponding to different parameters fail to commute, no single measurement can simultaneously saturate the QCRB for all parameters. Notably, sloppiness (from encoding dynamics) and incompatibility (from measurement) represent distinct sources of estimation uncertainty. Understanding this trade-off is critical for advancing multiparameter quantum metrology and achieving practical precision enhancements. 

Incompatibility in multiparameter quantum estimation has been investigated in different systems \cite{zhu2015,heinosaari2016,ragy2016,candeloro2024dimension}, and sloppiness have been discussed in several metrological scenarios \cite{brown2003statistical,brown2004statistical,PhysRevLett.97.150601,machta2013parameter,PRXQuantum.2.020308,Goldberg21,Yang23,Frigerio2024,PhysRevA.111.012414,adani2024,cavazzoni2024}, but little emphasis has been devoted to the link between sloppiness and incompatibility.  In this paper, we address a two-parameter qubit statistical model with tunable sloppiness and present an adjustable scrambling operation to investigate how the correlations between the parameters and the incompatibility between the SLDs influence the precision bounds.

This work is structured as follows. Section 2 introduces the fundamentals of quantum multiparameter estimation. In Section 3, to explore the interplay between sloppiness and incompatibility, we present a two-parameter qubit estimation model that incorporates an information scrambling operation. This operation allows us to tune the correlations between parameters and, consequently, the sloppiness of the model. We also analyze the trade-off between sloppiness and incompatibility. In Section 4, we optimize the relevant quantum Cramér-Rao bounds (QCRBs) and examine the role of sloppiness in achieving these bounds. Finally, Section 5 concludes the paper with a summary of our findings. Details of calculations are provided in the Appendices.

\section{Multiparameter quantum estimation: precision, sloppiness and incompatibility}
In this section, we provide the theoretical framework, definitions and metrics used throughout the paper. We consider finite dimensional systems  and a family of 
quantum states $\rho_{\bm \lambda}$ encoding the values of $d$ real parameters, denoted 
as a vector $\bm\lambda=(\lambda_0, \lambda_1,\ldots, \lambda_d)^T$. If we perform a positive operator-valued measurement (POVM) $\mathbf \Pi$ with elements $\left\{\Pi_k\right\}$ satisfying $\Pi_k\geq0$ and $\sum_{k} \Pi_k=\mathbb{I}$, the measurement outcome $k$ is obtained with probability $p_{\bm \lambda}(k) = \Tr \left[\rho_{\bm \lambda}\Pi_k\right]$. The estimator function based on the result is denoted as $\hat{\bm \lambda}(k)$. The performance of the estimator is assessed by the covariance matrix $\mathbf V(\hat{\bm \lambda})$ with elements
\begin{align*}
V_{\mu\nu} =&\sum_k p_{\bm \lambda}(k)[\hat{ \lambda}_\mu(k)-E_k(\hat \lambda_\nu)][\hat{ \lambda}_\nu(k)-E_k(\hat \lambda_\nu)].
\end{align*}
where $E_k(\hat \lambda_\mu)$ is the expectation value of $\hat{\lambda}_\mu$ over the probability distribution $p_{\bm \lambda}(k)$. 

In classical multiparameter estimation, when the estimators satisfying the locally unbiased conditions:
\begin{equation*}
\bm E_\nu(\hat {\bm \lambda})=\hat{\bm \lambda}\, \qquad \partial_\mu E_k(\hat \lambda_\nu)=\delta_\mu^\nu,
\end{equation*}
where $\partial_\mu=\frac{\partial}{\partial \lambda_\mu}$,
then the CRB \cite{cramer1999} holds
\begin{equation*}
\mathbf V(\hat{\bm \lambda})\geq \frac{1}{M \mathbf F}\,,
\end{equation*}
where $M$ is the number of repeated measurements and $\mathbf F$ is the FI matrix with elements defined by
\begin{equation*}
F_{\mu\nu}=\sum_k p_{\bm \lambda}(k)\,\partial_\mu \log p_{\bm \lambda}(k)\, \partial_\nu \log p_{\bm \lambda}(k) = \sum_k \frac{\partial_\mu  p_{\bm \lambda}(k)\, \partial_\nu  p_{\bm \lambda}(k)}{p_{\bm \lambda}(k)}
\,.
\end{equation*}
The CRB can be saturated in the asymptotic limit of an infinite number of repeated experiments using Bayesian or maximum likelihood estimators \cite{kay1993}. 

Due to the non-commutativity of the operators on $\mathcal{H}$, the quantum analogue of the FI cannot be uniquely introduced. In fact, there exist several different definitions of quantum Fisher information. The most celebrated and useful approaches are based on the so-called 
symmetric logarithmic derivative (SLD) operators $L_\mu^S$ \cite{helstrom1967} and right logarithmic derivative (RLD) operators $L_\mu ^R$ \cite{Yuen1973,FujiwaraA1994}, defined as follows
\begin{align*}
\partial_\mu\rho_{\bm \lambda}&=\frac{L_\mu^S\rho_{\bm \lambda}+\rho_{\bm \lambda} L_\mu^S}{2}\,, \\\partial_\mu \rho_{\bm \lambda}&=\rho_{\bm \lambda} L_\mu ^R.
\end{align*}
We denote the corresponding SLD and RLD quantum Fisher information matrices (QFIM) as $\mathbf{Q}$ and $\mathbf{J}$, respectively, with  elements
\begin{align*}
Q_{\mu \nu}&=\frac{1}{2}\Tr\left[\rho_{\bm \lambda}\{L_\mu^S,L_\nu^S\}\right]\,, \\
J_{\mu \nu}&=\Tr\left[\rho_{\bm \lambda} L_\mu^R L_\nu^{R\dagger}\right]\,.
\end{align*}
For pure statistical models, $\rho_{\bm \lambda}=|\psi_{\bm \lambda}\rangle\langle\psi_{\bm \lambda}|$ we have 
\begin{align*}
Q_{\mu\nu} & = 4\,\textrm{Re}\big(\ip{\partial_\mu\psi_\lambda}{\partial_\nu\psi_\lambda}-\ip{\partial_\mu\psi_\lambda}{\psi_\lambda}\ip{\psi_\lambda}{\partial_\nu\psi_\lambda}\big),\\
Q_{\nu\mu} & = Q_{\mu\nu}\,,
\end{align*}
where $\partial_k\equiv\partial_{\lambda_k}$.
\subsection{Symmetric and right quantum Cram$\acute{e}$r-Rao Bounds}  
Using the above SLD and RLD QFIMs, $\mathbf{Q}$ and $\mathbf{J}$, matrix inequalities for the covariance matrix of any set of locally unbiased estimators may be established. Then, in order to obtain a scalar bound and to tailor the optimization of precision according to the different applications, a weight matrix $\mathbf W$ (a positive, real $d\times d$ matrix) may be introduced. The corresponding symmetric and right scalar bounds read as follows: 
\begin{align*}
C_S[W,{\hat{\bm \lambda}}] & =\frac1M\,\Tr\left[\mathbf W\, \mathbf{Q}^{-1}\right],\\
C_R[W,{\hat{\bm \lambda}}]& = \frac1M\, \Tr\left[\mathbf W\, \textrm{Re}(\mathbf{J}^{-1})\right]+\Tr \left[\left|{\mathbf W\, \textrm{Im}(\mathbf{J}^{-1}) }\right|\right]\,,
\end{align*}
where $|A|=\sqrt{A^\dag A}$ and $\textrm{Re}(A)$ and $\textrm{Im}(A)$ denote 
the real and imaginary parts of the complex-valued matrix $A$, respectively.

\subsection{Holevo and Nagaoka Cram$\acute{e}$r-Rao Bounds} 
If the SLDs do not commute, it may happen that measurements that are optimal for different 
parameters are incompatible, making the symmetric and right QCRB, as well as their scalar counterparts, not achievable. An achievable scalar bound has been derived by Holevo \cite{holevo2011}:
\begin{equation*}
C_H[\mathbf W,{\hat{\bm \lambda}}]
=\min_{\mathbf X \in \mathcal X} \Big\{\Tr\left[\mathbf W\, \textrm{Re}\left(\mathbf Z[\mathbf X]\right)\right]+\Tr \left[\left|\mathbf W\, \textrm{Im}\left(\mathbf Z[\mathbf X]\right)\right|\right]\Big\},
\end{equation*}
where 
the Hermitian $d\times d$ matrix $\mathbf Z$ is defined via its elements $\mathbf Z_{\mu\nu}[\mathbf X] = \Tr \left[\rho_{\bm \lambda}X_\mu X_\nu\right]$ with the collection of Hermitian operators $\mathbf X$ belonging to the set $\mathcal X=\{\mathbf X =(X_1, \ldots, X_d)| \Tr[(\partial _\mu \rho_{\bm \lambda})X_\nu]=\delta_\nu^\mu\}$. 
 It has been proven that Holevo CRB $C_H[\mathbf W,{\hat{\bm \lambda}}]$ becomes attainable by performing a collective measurement on an asympotically large number of copies of the state $\rho_{\bm \lambda}^{\bigotimes n}$ with $n\rightarrow\infty$. As such, it is typically regarded as the most fundamental scalar bound for multiparameter quantum estimation.  
 
Nagaoka \cite{Nagaoka2005} introduced a more informative bound, denoted as $C_N[\mathbf W,{\hat{\bm \lambda}}]$, which is particularly valuable for practical experimental measurements. While not as theoretically tight as the Holevo bound $C_H[\mathbf W,{\hat{\bm \lambda}}]$, it can be practically achieved using separable measurement strategies, making it more feasible to attain in real-world applications. The bound is defined as: 
  \begin{equation}
C_N[\mathbf W,{\hat{\bm \lambda}}]= \min_{\mathbf{\Pi} }\Big\{\Tr\left[\mathbf W \mathbf F^{-1}\right] \Big\},
\end{equation}
where the minimization is performed over all possible single-system ($\equiv$ non collective) POVMs $\mathbf{\Pi}$. As for the other bounds, the optimal POVM generally depends on the true value of the parameters $\bm \lambda$.
 \subsection{Sloppiness and stiffness}
A quantum statistical model is termed sloppy if the QFIM is singular, 
i.e. $\det [\mathbf Q]=0$. This means that the true parameters describing the 
system are $m < n$ combinations of the original parameters ${\lambda_1, 
\lambda_2, \ldots, \lambda_n}$.  The eigenvalues of $\mathbf Q$ quantify the sensitivity of the probe state to perturbations along orthogonal parameter directions. A small eigenvalue implies that the state of the probe is insensitive to changes in the corresponding parameter direction, i.e., that combination of parameters is poorly encoded. The degree of sloppiness may be thus quantified by the  determinant of QFIM:
 \begin{equation}
s:= \frac{1}{\det [\mathbf Q]} = \det [\mathbf Q^{-1}]\,,
\end{equation}
which measures how strongly the system depends on a combination of the components of $\bf \lambda$ rather than on its individual components.  A large sloppiness indicates that the model’s sensitivity to parameter variations is highly anisotropic. A model with low sloppiness is referred to as a stiff model.
\subsection{Compatibility and incompatibility}
Due to the non-commutativity of the SLDs associated to different parameters, in a multiparameter scenario QCRB cannot always be achieved. Based on the quantum local asymptotic normality\cite{hayashi2008,kahn2009,Yamagata2013}, it has been shown that 
the multiparameter SLD-QCRB is attainable if and only if the weak compatibility condition 
is satisfied \cite{ragy2016}, defined by
\begin{equation}
\Tr \left[\rho_{\bm \lambda} [L_\mu^S, L_\nu^S]\right]=0.\label{WCC}
\end{equation}
The incompatibility matrix $\mathbf D$, also known as mean Uhlmann curvature (MUC), is the antisymmetri matrix defined by 
\begin{equation}
{D}_{\mu\nu}:=\frac{1}{2i}\Tr \left[ \rho_{\bm \lambda} [L_\mu^S, L_\nu^S]\right]\,,
\end{equation}
and is useful to quantify the incompatibility between the pair of parameters 
$\lambda_\mu$ and $\lambda_\nu$.  For a two-parameter pure state model we have
\begin{align*}  
D_{11} & =D_{22}   =0, \\
D_{12} & =-D_{21}  = 4\,\textrm{Im}\left(\ip{\partial_1\psi_\lambda}{\partial_2\psi_\lambda}-\ip{\partial_1\psi_\lambda}{\psi_\lambda}\ip{\psi_\lambda}{\partial_2\psi_\lambda}\right).
\end{align*}
A measure of the compatibility of the model is given by 
\begin{equation*}
c := \frac{2}{ \Tr\left[\mathbf D^{\dagger}\mathbf D\right]}\,,
\end{equation*}
which is based on the antisymmetric nature of the MUC. This measure is non-negative for all pairs of SLD operators and approaches infinity when all SLD operators are commuting with each other.  Incompatibility is affected by the parameters' component, and thus $c$ is not invariant 
under reparametrization. However, incompatibility is covariant in some natural scenarios in which it should not be possible to generate incompatibility, such as unitary evolution. 
In our 2-parameter qubit estimation model, the incompatibility can expressed as
\begin{equation}
c:= \frac1{\det [\mathbf D]} = \det [{\mathbf D}^{-1}] = - \frac{1}{D_{12}^2}\,.
\end{equation}
\subsection{Relationship between different bounds} 
 The relationship between the different bounds has been extensively explored widely, providing 
 a classification of quantum statistical models. Specifically, the models can be categorized 
 into four types:
 \begin{itemize}
\item {\bf Classical} When the quantum state $\rho_{\bm \lambda}$ can be expressed in a diagonal form with parameters $\bm \lambda$ and a $\bm \lambda$-independent unitary $U$,such that $\rho_{\bm \lambda}=U\Lambda_{\bm \lambda}U^{\dagger}$, it can be regarded as a classical quantum statistical model. In this scenario, the FI matrix $\mathbf F$, the SLD QFIM $\mathbf Q$, and the RLD QFIM $\mathbf J$ are all identical. Consequently, we obtain $C_S[\mathbf W,{\hat{\bm \lambda}}]=C_R[\mathbf W,{\hat{\bm \lambda}}]=C_H[\mathbf W,{\hat{\bm \lambda}}]=C_N[\mathbf W,{\hat{\bm \lambda}}]$.
\item {\bf Quasi-classical} A quantum statistical model is called quasi-classical if all SLD operators commute with each other for every parameter. Thus, the equality $C_S[\mathbf W,{\hat{\bm \lambda}}]=C_H[\mathbf W,{\hat{\bm \lambda}}]$ holds.
\item {\bf Asymptotically classical} In asymptotically classical quantum statistical models, all SLD operators satisfy the weak compatibility condition defined in Eq.(\ref{WCC}), and $C_S[\mathbf W,{\hat{\bm \lambda}}]$ is equal to $C_H[\mathbf W,{\hat{\bm \lambda}}]$.
\item {\bf D-invariant} When a quantum statistical model is D-invariant, we have \cite{FujiwaraA1994,Suzuki2016}
\begin{equation*}
C_R[\mathbf W,{\hat{\bm \lambda}}]=C_H[\mathbf W,{\hat{\bm \lambda}}]=C_S[\mathbf W,{\hat{\bm \lambda}}]+ \left|\sqrt{\mathbf {W}}\mathbf Q^{-1}\mathbf U \mathbf Q^{-1}\sqrt{\mathbf{W}}\right|\,.
\end{equation*}
This equality illustrates that the RLD bound is achievable by performing a collective measurement on an asymptotically large number of copies.
\end{itemize}

In \cite{carollo2019} a measure has been suggest to quantify the amount of incompatibility 
(somehow referred to as the {\em quantumness} of the model \cite{razavian2020quantumness})
\begin{equation}
R:= ||i\mathbf{Q}^{-1} \mathbf D ||_{\infty},
\end{equation}
where $ ||A||_{\infty}$ denotes the largest eigenvalue of the matrix $A$.  It has been also 
shown that $R$ is useful in upper bounding the Holevo bound in terms of the SLD-bound, as follows:
\begin{align}
		C_S[\mathbf W,{\hat{\bm \lambda}}]\leq C_H[\mathbf W,{\hat{\bm \lambda}}]\leq C_N[\mathbf W,{\hat{\bm \lambda}}]\leq (1+R)C_S[\mathbf W,{\hat{\bm \lambda}}]\leq 2C_S[\mathbf W,{\hat{\bm \lambda}}]\,.\label{ine}
\end{align}
In the pure-state limit, the Holevo bound is equivalent to the RLD CR bound \cite{Suzuki2016}. When the number of parameters to be estimated is $n = 2$,  we have the relation 
	\begin{equation}\label{defR}
	R=\sqrt{\frac{\det\left[\mathbf D\right]}{\det\left[ \mathbf Q\right] }}=\sqrt{\frac{s}{c}},
	\end{equation}
	where $s$ represents the sloppiness and $c$ is the compatibility of the system.
For our qubit multiparameter model,  Eq.(\ref{ine}) can be further rewritten as
\begin{equation}
C_S[\mathbf W,{\hat{\bm \lambda}}]\leq C_H[\mathbf W,{\hat{\bm \lambda}}]=C_N[\mathbf W,{\hat{\bm \lambda}}]\leq C_S[\mathbf W,{\hat{\bm \lambda}}]\left(1+\sqrt{\frac{s}{c}}\right).\label{e2}
\end{equation}
\section{Information scrambling, precision, and the sloppiness-incompatibility trade-off}
\begin{figure}[!ht]
	\centering
	\includegraphics[width=0.8\textwidth]{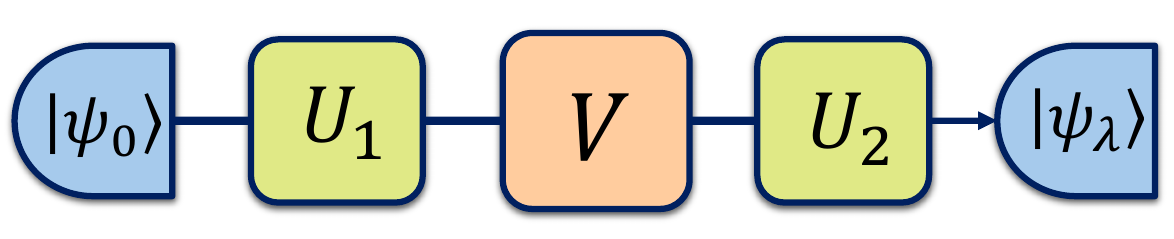}
	\caption{The scrambling model considered in this paper. The model parameters $\lambda_1$ and $\lambda_2$ are encoded via the 
unitary operations $U_1$ and $U_2$, which represent rotation along the z-axis of the Bloch sphere. To remove sloppiness and adjust correlations between the encoded parameters, we introduce a scrambling operation, represented by the intermediate rotation $V$.}\label{f1}
\end{figure}

To systematically investigate the interplay between sloppiness and incompatibility, we consider
the two-parameter qubit model illustrated in Fig.\ref{f1}. By introducing a tunable scrambling operation during parameter encoding, we control correlations between parameters and quantify their impact on estimation precision. Given the convexity of QFI \cite{PhysRevA.91.042104}, we consider a pure probe state $\ket{\psi_0}$ defined as
\begin{equation*}
\ket{\psi_0}=\cos \frac{\alpha}{2} \ket{0}+ e^{i\beta}\sin  \frac{\alpha}{2} \ket{1}\,.
\end{equation*}
The model parameters $\lambda_1$ and $\lambda_2$ are encoded via the 
unitary operations $U_1$ and $U_2$, which represent rotation
along the z-axis of the Bloch sphere, and are given by 
\begin{equation*}
 U_k=e^{-i\sigma_3\lambda_k},
  \end{equation*}
where $\sigma_3$ is the Pauli-$Z$ matrix.  These rotations imprint $\lambda_1$ and $\lambda_2$ onto the probe state's phase.

If nothing is done between the two unitaries, the output state depends only on the sum of the two parameters (and not on the difference), the QFIM is thus singular and the model is sloppy. To remove sloppiness in a tunable way, and adjust correlations between the encoded parameters, we introduce an intermediate rotation $V$ between $U_1$ and $U_2$: 
 \begin{equation*}
 V=e^{-i\gamma\vec{\sigma}\cdot \vec{n}}\,, \qquad \vec{n}=(\cos\phi \sin\theta, \sin \phi \sin\theta, \cos \theta).
 \end{equation*}
 
 Here, $\gamma$ controls the rotation strength, $\theta$ and $\phi$ define the rotation axis $\vec{n}$, and $\vec{\sigma}$ is the vector of Pauli matrices. This operation dynamically mixes the parameters, introducing correlations that govern sloppiness. The final state after encoding becomes
\begin{align*}
\fl \ket{\psi_\lambda}
& =U_2VU_1\ket{\psi_0} \\ &=\left( {\begin{array}{c}
	e^{-i (\lambda_1+\lambda_2)} \cos\frac{\alpha}{2} (\cos \gamma-i\sin \gamma \cos \theta)-ie^{i (\lambda_1-\lambda_2+\beta -\phi )} \sin\frac{\alpha}{2} \sin \gamma \sin \theta  \\
	-i e^{-i(\lambda_1-\lambda_2 -\phi)}\cos \frac{\alpha}{2} \sin  \gamma \sin \theta + e^{i (\lambda_1+\lambda_2+\beta )} \sin\frac{\alpha}{2} (\cos \gamma+i\sin \gamma \cos \theta)
\label{s1}  \end{array} } \right)\,. \notag
	\end{align*}
	
Explicitly, this state depends on 7 parameters. $\alpha$ and $\beta$ are probe state initialization's parameters. $\alpha$ balances the superposition weights of $\ket{0}$ and  $\ket{1}$. $\beta$ is the initial phase, influencing interference effects during the parameter encoding. $\lambda_1$ and $\lambda_2$ are encoding parameters, $\gamma$, $\theta$, and $\phi$ are scrambling parameters. 

The rotation angle $\gamma$ partly governs the strength of parameter mixing. When $\gamma =0$, $V={\mathbb I}$, and the parameters perfectly correlated as they combine into a single effective parameter $\lambda_1+\lambda_2$. This results in maximum sloppiness ($\det \mathbf Q \rightarrow 0$) because only a single function of the parameters can be estimated. Larger $\gamma$ may decrease coupling between $\lambda_1$ and $\lambda_2$, allowing $U_1$ and $U_2$ to imprint independent information the quantum state. The angle $\theta$ determines the alignment of rotation axis relative to the $Z$-axis, and $\phi$ controls the azimuthal orientation of rotation axis, introducing phase-dependent correlations. The scrambling rotation $V$ couples $\lambda_1$ and $\lambda_2$, enabling control over the model's sloppiness and the non-commutativity of their associated SLDs.

Explicit calculations yield:
 \begin{align*}
 		Q_{11}&=4\sin ^2 \alpha, \\
 		Q_{12}&=4(X \sin ^2 \alpha  - Y \sin 2\alpha ),\\
 		Q_{22}&=4[1-(X \cos \alpha +2 Y\sin \alpha)^2],
 	\end{align*}
where 
\begin{align*}
X&=\cos ^2 \gamma +\sin ^2 \gamma \cos 2\theta, \\
Y&=\sin \gamma \sin \theta (\sin f \cos \gamma+ \sin \gamma \cos \theta \cos f ), \\ 
f&=2\lambda_1+\beta -\phi\,.
\end{align*}

The measurement incompatibility matrix $\mathbf D$ has elements:
\begin{eqnarray*}
 		D_{11}&=D_{22}=0, \\
 		D_{12}&=-D_{21}=-8Z\sin \alpha \,,
 	\end{eqnarray*}
where 
$$Z=\sin \gamma \sin \theta (\cos f \cos \gamma- \sin \gamma \cos \theta \sin f )\,.$$
From these results, we are to calculate the sloppiness and compatibility measures as follows
  \begin{align}
 1/s & = \det[\mathbf Q]=16\sin^2 \alpha[1-X^2-4Y^2]=64\,Z^2\,\sin ^2 \alpha ,\\
  1/c  &= \det[\mathbf{D}]=64\,Z^2\,\sin ^2 \alpha\, , \label{d}
\end{align}
uncovering the  fundamental trade-off:
  \begin{equation}
  s=c\,. \label{td}
    \end{equation}
This equality quantifies the competition between parameter distinguishability and compatibility 
in our two-parameter qubit model. In other words, sloppiness and incompatibility cannot be minimized simultaneously. More detailed derivations are provided in Appendix 1. Notice that Eq. (\ref{td}), together with Eq. (\ref{defR}), is consistent with the observation \cite{candeloro2024dimension} that when the number of parameters is equal to the dimension of the probe, we have maximum quantumness of the model, i.e., $R=1$.
\section{The ultimate bounds to precision}
The interplay between sloppiness and incompatibility in multiparamater quantum estimation is not merely a theoretical relation, it also informs strategies for optimizing precision. Here, we explore this optimization for a two-parameter qubit estimation system, revealing how suitable parameter encoding and measurement design achieve these goals. 
\subsection{Hierachy of Quantum Cram$\acute{e}$r-Rao Bounds}
In our two-parameter qubit model, we assume that the two parameters are equally relevant,
and thus set $\mathbf W={\mathbb I}$. We start by noticing that for a positive definite 
matrix, we have the property $\frac1n \Tr[A^{-1}] \geq \det[A]^{-1/n}$. 
For $n=2$, we obtain
	\begin{equation*}
	\frac{2}{\Tr[\mathbf {Q}^{-1}]}\leq \sqrt{\det[\mathbf Q]}.
	\end{equation*}
Therefore, in our qubit model, the different quantum Cram\'er-Rao bounds with $\mathbf W={\mathbb I}$ satisfy
\begin{equation}
	 2\sqrt{s}\leq \Tr(\mathbf {Q}^{-1})=C_S\leq C_H=C_N\leq C_S\left(1+\sqrt{\frac{s}{c}}\right)=2\,C_S\,,\label{e3}
\end{equation}
where the dependence of the bounds on the weight matrix and the parameters has been dropped 
since the weight matrix is set to the identity and the model is unitary (i.e., the bounds do not depends on the parameters). The SLD bound is not achievable, while the larger Holevo and Nagaoka bounds are achievable, and equivalent in this model. The upper bound remains tied to the SLD bound.
\subsection{Minimization of SLD bound}
From Eq.(\ref{e3}), to minimize $C_S$, we need to minimize the sloppiness, i.e., maximize $\det [\mathbf Q]$. Given Eq. (\ref{d}), the task is linked to maximizing the quantitity 
$Z^2$. Two parameter regimes achieve this: 
$\textrm{Case (a)}$: $\gamma=f=\frac{\pi}{2}, \;\theta =\frac{\pi}{4},$
$\textrm{Case (b)}$: $\gamma=\frac{\pi}{4},\,f=0,\, \theta=\frac{\pi}{2}.$
This result underscores the necessity of adjusting $\gamma, \theta, f$ to decouple $\lambda_1$ and $\lambda_2$. Detailed calculations are provided in Appendix 2. Both cases yield $Z^2=1/4$, saturating the upper limit of incompatibility. When combined with an optimal probe state ($\alpha = \pi/2$), sloppiness reaches its minimum ($s=1/16$), and the SLD bound $C_S$ attains its lowest value:
\begin{equation*}
C_S=\frac{1}{2}.
\end{equation*}
In our two-parameter qubit model, we achieve  SLD precision by reducing sloppiness. It means 
that theoretical maximum precision can only be reached when sloppiness is minimized.
\subsection{Minimizing the Holevo and Nagaoka bounds} 
For a two-parameter pure qubit estimation model, the Holevo bound 
$C_H[\mathbf W,{\hat{\bm \lambda}}]$ is given by
\begin{equation*}
	C_H[\mathbf W,{\hat{\bm \lambda}}] = C_R[\mathbf W,{\hat{\bm \lambda}}]= C_H[\mathbf W,{\hat{\bm \lambda}}] + \frac{\sqrt{\det [\mathbf W]}}{\det[ \mathbf Q]} |\Tr\left[\rho_{\bm \lambda} [L^S_1, L^S_2]\right]|. \label{ch}
\end{equation*}
Similarly, the Nagaoka bound $C_N[\mathbf W,{\hat{\bm \lambda}}]$ for 
a two-parameter qubit model is given by
\begin{equation*}
	C_N[\mathbf W,{\hat{\bm \lambda}}]= C_S[\mathbf W,{\hat{\bm \lambda}}]+ \frac{\sqrt{\det [\mathbf W]}}{\det [\mathbf Q]} \textrm{Tr}\left[\left|\rho_{\bm \lambda} [L^S_1, L^S_2]\right|\right]\,. \label{cn}
\end{equation*}
It follows directly that the Holevo and the Nagaoka bounds are identical for a two-parameter pure qubit \cite{Suzuki2016}, simplifying to
\begin{equation*}
	C_H[\mathbf W,{\hat{\bm \lambda}}] = C_N[\mathbf W,{\hat{\bm \lambda}}]= \Tr[\mathbf W\mathbf Q^{-1}]+ 2\sqrt{\det[\mathbf W\mathbf Q^{-1}]}. \label{ceq}
\end{equation*}
In our model, the Holevo bound, RLD bound and Nagaoka bounds are identical and given  
by
\begin{equation*}
	C_H= C_R=C_N= C_S + 2\sqrt{s}\geq4\sqrt{s}.
\end{equation*}
We find that all these bounds are expressed as lower bound in terms of sloppiness. When sloppiness achieves minimum, all quantum CR bounds reach their minima simultaneously, reducing to:
\begin{equation*}
C_H= C_R=C_N = 1.
\end{equation*}
This, together with the minimal condition of the SLD QCRB, indicates that  maximum estimation precision can be achieved by minimizing sloppiness.

Holevo bound represents the asymptotically achievable precision limit with collective measurements,
performed on all the available copies, say $M$, of the state encoding the parameters. Upon performing separate measurements (each one performed on one of the $M$ copies) one may 
reach the Nagaoka bound, which is usually larger than the Holevo bound. However in our model, these two bounds are equivalent, also indicating that our model is D-invariant. Notice that in our model $R=1$, and $C_H=2\,C_S$, i.e. the quantumness $R$ quantifies exactly the additional uncertainty due to the incompatibility between the two SLDs.
\subsection{Optimization of bounds for separate and successive measurements} 
Successive measurements involve estimating parameters sequentially, rather than estimating 
them jointly. Having at disposal $M$ repeated preparations of the system, we assume to 
devote $M/2$ of them to estimate solely $\lambda_1$ (assuming $\lambda_2$ unknown) and the remaining $M/2$ preparations to estimate $\lambda_2$ (assuming $\lambda_1$ known from the first step). Of course the role of the two parameters may be exchanged, and we thus have two strategies of this kind.

The (saturable) precision bound on the estimation of $\lambda_1$ from the first step is obtained from the SLD-QCRB by choosing a weight matrix of form $\mathbf W=\hbox{Diag}(1,0)$, leading to
\begin{equation*}
	\textrm{Var}\lambda_1 \geq \frac{2[\mathbf Q^{-1}]_{11}}{M},
\end{equation*}
where $[X]_{ij}$ indicates the elements of the  matrix $X$. In the second step, 
$\lambda_1$ is known, and the (achievable) bound to precision in the estimation 
of $\lambda_2$ is given by the single-parameter QCRB
\begin{equation*}
	\textrm{Var}\lambda_2 \geq \frac{2}{M Q_{22}}.
\end{equation*}
The total variance for this estimation strategy is thus bounded by: 
\begin{equation*} 
\textrm{Var}\lambda_1 + \textrm{Var}\lambda_2 \geq 
\frac{2}{M}\bigg([\mathbf Q^{-1}]_{11} + \frac{1}{Q_{22}}\bigg) \equiv \frac1M K_1,
\end{equation*}
where
\begin{equation*}
	K_1 = 2\bigg([\mathbf Q^{-1}]_{11}+ \frac{1}{Q_{22}}\bigg).
\end{equation*}
Similarly, reversing the role of the two parameters, we have $\textrm{Var}\lambda_1 + \textrm{Var}\lambda_2 \geq  K_2/M$, where
\begin{equation*}
	K_2 = 2\bigg([\mathbf Q^{-1}]_{22}+ \frac{1}{Q_{11}}\bigg).
\end{equation*}

For our model, we have
	\begin{eqnarray*}
K_1=2\bigg(s\,Q_{22}+\frac{1}{Q_{22}}\bigg)\geq 2\times 2\sqrt {s\,Q_{22}\times\frac{1}{Q_{22}}}=4\sqrt{s},\\
K_2=2\bigg(s\,Q_{11}+\frac{1}{Q_{11}}\bigg)\geq 2\times 2\sqrt {s\,Q_{11}\times\frac{1}{ Q_{11}}}=4\sqrt{s}.
\end{eqnarray*}
The equalities hold if and only if $s$ achieves its minimum, i.e., the minimum values of $K_1$ and  $K_2$ are equal.  Notice that if, instead of dividing the total number of repeated preparations equally between the two estimation procedures, we had chosen an asymmetric allocation, say $ M_1 = \gamma M$ measurements for estimating $ \lambda_1 $ and $M_2 = (1 - \gamma) M$ for estimating $\lambda_2 $ (or vice versa), the bounds $K_1$ and $K_2$ would have been larger, specifically $ K_j\geq 2\sqrt{s}/\sqrt{\gamma (1 - \gamma)} $, $j=1,2$. The choice $ \gamma = 1/2 $ is therefore optimal.
Overall, the relation among the precision bounds is
\begin{equation*}
	C_H= C_R=C_N=4\sqrt{s}\leq K_1 = K_2,
\end{equation*}
where $K_1=K_2=4\sqrt{s}$ if and only if the sloppiness is minimal. This equality demonstrates that even when the incompatibility is maximal, sequential measurements can still  achieve optimal precision. Sloppiness sets the lower bounds for both the Holevo/Nagaoka bound as well as the bound to precision achievable by successive measurements. The latter bound is identical to the Holevo bound if and only if the sloppiness is minimal, which offers a practical advantage in experiments. From another perspective, incompatibility can be used to improve precision in quantum multiparameter models. 
\section{Conclusions}
In this work, we have investigated a two-parameter qubit statistical model with tunable sloppiness to explore the interplay between precision, sloppiness, and incompatibility. By introducing an adjustable scrambling operation during parameter encoding, we have demonstrated how parameter correlations and measurement incompatibility jointly influence the precision bounds.

First, we identified a fundamental trade-off between sloppiness and incompatibility, characterized by the equality $s = c$. This result highlights the impossibility of simultaneously minimizing both sloppiness and incompatibility, revealing a key constraint in multiparameter quantum estimation in 
qubit systems. Second, we derived the conditions for optimizing quantum Cram\'er-Rao bounds. By tuning the parameters of the encoding strategy, we maximized the determinant of the quantum Fisher information matrix, thereby minimizing sloppiness.

In our system, the Holevo, Nagaoka, and right logarithmic derivative precision bounds for joint parameter estimation are equivalent and saturable using non-collective measurements. We also compared the performance of joint estimation strategies to those involving successive separate estimation steps, demonstrating that the former can achieve ultimate precision when sloppiness is minimized. Beyond its fundamental significance, this finding offers practical advantages for experimental implementations. Furthermore, our analysis revealed that the minimal achievable precision bounds directly connect sloppiness to the ultimate metrological performance. This underscores the importance of designing probe states and encoding dynamics that emphasize parameter correlations while balancing the effects of non-commutative measurements.

Our results provide new insights into the relationship between sloppiness and incompatibility in two-parameter qubit estimation systems. In future work, we aim to extend this framework to higher-dimensional systems, where careful preparation of the probe state may eliminate incompatibility \cite{candeloro2024dimension} and render the symmetric Cramér-Rao bound achievable. This could lead to fundamentally different trade-offs between sloppiness, precision, and incompatibility compared to the qubit case.

\section*{Acknowledgements}
This work has been done under the auspices of GNFM-INdAM and has been partially supported by MUR through Project No. PRIN22-2022T25TR3-RISQUE. MGAP thanks Francesco Albarelli, Chiranjib Mukhopadhyay, Abolfazl Bayat, Sara Dornetti, Alessandro Ferraro, Berihu Teklu, Victor Montenegro, Simone Cavazzoni, and Stefano Olivares for useful discussions. JH thanks Shuangshuang Fu for useful discussions.
\appendix

\section{Quantum Fisher information matrix and mean Uhlmann curvature}
We first calculate the output state
\begin{align*}
\fl \ket{\psi_\lambda}& =U_2VU_1\ket{\psi_0} \\ & =\left( {\begin{array}{c}
	e^{-i (\lambda_1+\lambda_2)} \cos\frac{\alpha}{2} (\cos \gamma-i\sin \gamma \cos \theta)-ie^{i (\lambda_1-\lambda_2+\beta -\phi )} \sin\frac{\alpha}{2} \sin \gamma \sin \theta  \\
	-i e^{-i(\lambda_1-\lambda_2 -\phi)}\cos \frac{\alpha}{2} \sin  \gamma \sin \theta + e^{i (\lambda_1+\lambda_2+\beta )} \sin\frac{\alpha}{2} (\cos \gamma+i\sin \gamma \cos \theta) 
	 \end{array} } \right).
\end{align*}

The partial derivatives of the output state $\psi_\lambda$ with respect to $\lambda_1$ and $\lambda_2$, respectively, are given by
	\begin{eqnarray*} \label{P1}
	\fl \ket{\partial_{\lambda_1}\psi_\lambda }&= 
	\left( {\begin{array}{c}
			-i e^{-i (\lambda_1+\lambda_2)} \cos\frac{\alpha}{2} (\cos \gamma-i\sin \gamma \cos \theta)+e^{i (\lambda_1-\lambda_2+\beta -\phi )} \sin\frac{\alpha}{2} \sin \gamma \sin \theta  \\
			-e^{-i(\lambda_1-\lambda_2 -\phi)}\cos \frac{\alpha}{2} \sin  \gamma \sin \theta + i e^{i (\lambda_1+\lambda_2+\beta )} \sin\frac{\alpha}{2} (\cos \gamma+i\sin \gamma \cos \theta) \end{array} } \right),\\ 
		\fl \ket{\partial_{\lambda_2}\psi_\lambda }&= \left( {\begin{array}{c}
			-i e^{-i (\lambda_1+\lambda_2)} \cos\frac{\alpha}{2} (\cos \gamma-i\sin \gamma \cos \theta)	-e^{i (\lambda_1-\lambda_2+\beta -\phi )} \sin\frac{\alpha}{2} \sin \gamma \sin \theta)  \\
			e^{-i(\lambda_1-\lambda_2 -\phi)}\cos \frac{\alpha}{2} \sin  \gamma \sin \theta + i	e^{i (\lambda_1+\lambda_2+\beta )} \sin\frac{\alpha}{2} (\cos \gamma+i\sin \gamma \cos \theta)
		\end{array} } \right).  \label{P2}
	\end{eqnarray*}
which lead to
 \begin{eqnarray*}
\fl & 	\ip{\partial_{\lambda_1}\psi_\lambda}{\partial_{\lambda_1}\psi_\lambda}= \ip{\partial_{\lambda_2}\psi_\lambda}{\partial_{\lambda_2}\psi_\lambda}=1,\\
\fl& 	\ip{\partial_{\lambda_1}\psi_\lambda}{\partial_{\lambda_2}\psi_\lambda}=\cos^2\gamma+\sin^2\gamma\cos2\theta+2i\sin\alpha\sin \gamma\sin\theta (\sin \gamma\cos\theta\sin f-\cos f\cos r),\\
\fl& 	\ip{\partial_{\lambda_1}\psi_\lambda}{\psi_\lambda}=i\cos\alpha,\\
\fl& 	\ip{\partial_{\lambda_2}\psi_\lambda}{\psi_\lambda}= i\big[\!\cos\! \alpha(\cos^2\!\gamma+\sin^2\!\gamma\cos2\theta)\!+\!2\sin\! \alpha\sin\! \gamma \sin \!\theta (\sin \!f \cos\! \gamma+ \sin\! \gamma \cos\! \theta \cos \!f )\big],
 \end{eqnarray*}
where $f=2\lambda_1+\beta -\phi$. Then, we calculate the elements of QFIM $\mathbf Q$ and the incompatibility matrix $\mathbf D$. To make the computation easier, we define 
\begin{align*}
 		X&=\cos ^2 \gamma +\sin ^2 \gamma \cos 2\theta,\\
 		Y &=\sin \gamma \sin \theta (\sin f \cos \gamma+ \sin \gamma \cos \theta \cos f ),\\
 		Z&=\sin \gamma \sin \theta (\cos f \cos \gamma- \sin \gamma \cos \theta \sin f ).\\
 	 \end{align*}
 We obtain
 \begin{eqnarray*}
 \fl Q_{11}=4\sin ^2 \alpha, \\
 \fl Q_{12}=Q_{21}=4[\sin ^2 \alpha(\cos ^2 \gamma +\sin ^2 \gamma \cos 2\theta )-\sin 2\alpha  \sin \gamma \sin \theta (\sin f \cos \gamma+ \sin \gamma \cos \theta \cos f )] \\
\fl \,\,\,\,\,\,\,\,\,\,\,=4(X \sin ^2 \alpha  -Y \sin 2\alpha  ),\\
 \fl Q_{22}=4(1-[\cos \alpha(\cos ^2 \gamma +\sin ^2 \gamma \cos 2\theta )+2\sin \alpha \sin \gamma \sin \theta (\sin f \cos \gamma+ \sin \gamma \cos \theta \cos f )]^2) \\
 \fl\,\,\,\,\,\,\,\,\,\,\, =4[1-(X\cos \alpha +2Y\sin \alpha )^2],\\
 \fl	D_{11}=D_{22}=0, \\
 \fl	D_{12}=-D_{21}=4\textrm{Im}(\ip{\partial_{\lambda_1}\psi_\lambda}{\partial_{\lambda_2}\psi_\lambda}-\ip{\partial_{\lambda_1}\psi_\lambda}{\psi_\lambda}\ip{\psi_\lambda}{\partial_{\lambda_2}\psi_\lambda})\\
 \fl\,\,\,\,\,\,\,\,\,\,\,=-8\sin \alpha  \sin \gamma \sin \theta (\cos f \cos \gamma- \sin \gamma \cos \theta \sin f )=-8 X \sin \alpha\,.
 	\end{eqnarray*}
 The determinant of QFIM can be expressed as:
  \begin{equation*}
\det[\mathbf Q]= Q_{11}\times Q_{22}- Q_{12}\times Q_{21}=16\sin^2 \alpha[1-X^2-4Y^2]\,,
\end{equation*}
whereas the determinant of the incompatibility matrix is given by
  \begin{equation*}
 \det[\mathbf{D}]=0-D_{12}D_{21}=64\,Z^2\sin ^2 \alpha\, .
  \end{equation*}
More simplifications are required to clarify the connection between $\det[\mathbf Q]$ and $\det[\mathbf D]$.  Upon introducing the quantities $A=\sin \gamma \cos \theta, \, B=\sin \gamma \sin \theta$, we find that $	A^2+B^2=\sin ^2 \gamma$, so $\cos^2 \gamma=1-A^2-B^2$. Besides, $ Y^2+Z^2=\sin ^2\gamma \sin ^2\theta(\cos ^2 \gamma +\sin ^2 \gamma\cos ^2\theta)=B^2(1-A^2-B^2+A^2)=B^2(1-B^2)$ and $	X=\cos ^2 \gamma +\sin ^2 \gamma(\cos ^2\theta-\sin ^2\theta)=1-2B^2$, such that
\begin{eqnarray*}
\fl \det[\mathbf Q]=16\sin ^2 \alpha[1-(1-2B^2)^2-4F^2]	
	=64\sin ^2 \alpha[ B^2(1-B^2)-Y^2]=64\,Z^2\sin ^2 \alpha\,.
\end{eqnarray*}

\section{Maximum and minimum of $Z^2$}
To find the stationary points of $Z^2$, we calculate the partial derivatives of $Z^2$ with respect to $\gamma$, $\theta$, and $f$, and set them to zero:
  \begin{eqnarray*}
\fl \frac{	\partial Z^2}{\partial \gamma }=&2  \sin^2\theta \sin \gamma(\cos f \cos \gamma -\sin \gamma \cos \theta \sin f)\times[\cos f (\cos^2 \gamma -\sin ^2 \gamma )- 2 \sin \gamma \cos \gamma \cos \theta \sin f]=0,	\\
\fl	\frac{	\partial Z^2}{\partial \theta }=&2  \sin^2\gamma \sin \theta(\cos f \cos \gamma -\sin \gamma \cos \theta \sin f)\times[\cos f \cos \gamma \cos \theta-  \sin  \gamma \sin f(\cos^2 \theta -\sin ^2 \theta )]=0,	\\
\fl	\frac{	\partial Z^2}{\partial f }=&-2  \sin^2\gamma \sin ^2 \theta(\cos f \cos \gamma -\sin \gamma \cos \theta \sin f)(\cos \gamma \sin f+\sin  \gamma \cos \theta \cos f) =0.
\end{eqnarray*}
By analyzing these conditions, we find that stationary points occur in the following cases:
\begin{itemize}
\item[Case 1:] $\sin \gamma =0.$ In this case, $Z^2=0$, which leads to the minimum value of $d$.
\item[Case 2:] $\sin \theta =0.$ This case also leads to $Z^2=0$ and to the minimum value of $d$.
\item[Case 3:] $\cos \gamma \cos  f =\sin \gamma \cos \theta \sin f.$  This case also leads to $Z^2=0$ and to the minimum value of $d$.
\item[Case 4:] The system of equations:
 \begin{eqnarray*}
\fl	\cases{
		\cos f (\cos^2 \gamma -\sin ^2 \gamma )=2 \sin \gamma \cos \gamma \cos \theta \sin f,\\
		\sin  \gamma \sin f(\cos^2 \theta -\sin ^2 \theta )=\cos f \cos \gamma \cos \theta,\\
		\cos \gamma \sin f =-\sin  \gamma \cos \theta \cos f,
	}
\end{eqnarray*}
leads to two solutions that correspond to the maximum value of $Z^2$.
\end{itemize}
Those solutions are
\begin{itemize}
\item[Case (a):] $\gamma=f=\frac{\pi}{2}, \;\theta =\frac{\pi}{4},$
\item[Case (b):] $f=0, \theta=\frac{\pi}{2}, r=\frac{\pi}{4}.$
\end{itemize}
In both cases, the maximum value of $Z^2$ is $1/4$.
\section*{References}
\bibliography{CRB.bib}

\begin{thebibliography}{10}

\bibitem{demkowicz2020multi}
Rafa{\l} Demkowicz-Dobrza{\'n}ski, Wojciech G{\'o}recki, and
  M{\u{a}}d{\u{a}}lin Gu{\c{t}}{\u{a}}.
\newblock Multi-parameter estimation beyond quantum fisher information.
\newblock {\em Journal of Physics A: Mathematical and Theoretical},
  53(36):363001, 2020.

\bibitem{albarelli2020perspective}
Francesco Albarelli, Marco Barbieri, Marco~G Genoni, and Ilaria Gianani.
\newblock A perspective on multiparameter quantum metrology: From theoretical
  tools to applications in quantum imaging.
\newblock {\em Physics Letters A}, 384(12):126311, 2020.

\bibitem{liu2020quantum}
Jing Liu, Haidong Yuan, Xiao-Ming Lu, and Xiaoguang Wang.
\newblock Quantum fisher information matrix and multiparameter estimation.
\newblock {\em Journal of Physics A: Mathematical and Theoretical},
  53(2):023001, 2020.

\bibitem{razavian2020quantumness}
Sholeh Razavian, Matteo G.~A. Paris, and Marco~G Genoni.
\newblock On the quantumness of multiparameter estimation problems for qubit
  systems.
\newblock {\em Entropy}, 22(11):1197, 2020.

\bibitem{carollo2019}
Angelo Carollo, Bernardo Spagnolo, Alexander~A Dubkov, and Davide Valenti.
\newblock On quantumness in multi-parameter quantum estimation.
\newblock {\em Journal of Statistical Mechanics: Theory and Experiment},
  2019(9):094010, 2019.

\bibitem{PhysRevResearch.4.043057}
Mao Zhang, Huai-Ming Yu, Haidong Yuan, Xiaoguang Wang, Rafa\l{}
  Demkowicz-Dobrza\ifmmode~\acute{n}\else \'{n}\fi{}ski, and Jing Liu.
\newblock Quanestimation: An open-source toolkit for quantum parameter
  estimation.
\newblock {\em Phys. Rev. Res.}, 4:043057, Oct 2022.

\bibitem{sidhu20}
Jasminder~S. Sidhu and Pieter Kok.
\newblock Geometric perspective on quantum parameter estimation.
\newblock {\em AVS Quantum Science}, 2(1):014701, 02 2020.

\bibitem{Bonalda2019}
Daniele Bonalda, Luigi Seveso, and Matteo G.~A. Paris.
\newblock Quantum sensing of curvature.
\newblock {\em International Journal of Theoretical Physics}, 58(9):2914--2935,
  Sep 2019.

\bibitem{Danelli2025}
Maria Danelli and Matteo G.~A. Paris.
\newblock Are intrinsic decoherence models physical theories?
\newblock {\em Europhysics Letters}, 149(5):50001, feb 2025.

\bibitem{tsang2011}
Mankei Tsang, Howard~M Wiseman, and Carlton~M Caves.
\newblock Fundamental quantum limit to waveform estimation.
\newblock {\em Physical review letters}, 106(9):090401, 2011.

\bibitem{baumgratz2016}
Tillmann Baumgratz and Animesh Datta.
\newblock Quantum enhanced estimation of a multidimensional field.
\newblock {\em Physical review letters}, 116(3):030801, 2016.

\bibitem{Ansari2021}
Vahid Ansari, Benjamin Brecht, Jano Gil-Lopez, John~M. Donohue, Jaroslav
  \ifmmode \check{R}\else \v{R}\fi{}eh\'a\ifmmode~\check{c}\else \v{c}\fi{}ek,
  Zden\ifmmode \check{e}\else~\v{e}\fi{}k Hradil, Luis~L. S\'anchez-Soto, and
  Christine Silberhorn.
\newblock Achieving the ultimate quantum timing resolution.
\newblock {\em PRX Quantum}, 2:010301, Jan 2021.

\bibitem{Genovese2016}
Marco Genovese.
\newblock Real applications of quantum imaging.
\newblock {\em Journal of Optics}, 18(7):073002, jun 2016.

\bibitem{tsang2019}
Mankei Tsang.
\newblock Resolving starlight: a quantum perspective.
\newblock {\em Contemporary Physics}, 60(4):279--298, 2019.

\bibitem{ang2017}
Shan~Zheng Ang, Ranjith Nair, and Mankei Tsang.
\newblock Quantum limit for two-dimensional resolution of two incoherent
  optical point sources.
\newblock {\em Physical Review A}, 95(6):063847, 2017.

\bibitem{vrehavcek2017}
J~{\v{R}}eha{\v{c}}ek, Z~Hradil, B~Stoklasa, M~Pa{\'u}r, J~Grover, A~Krzic, and
  LL~S{\'a}nchez-Soto.
\newblock Multiparameter quantum metrology of incoherent point sources: Towards
  realistic superresolution.
\newblock {\em Physical Review A}, 96(6):062107, 2017.

\bibitem{vidrighin2014}
Mihai~D Vidrighin, Gaia Donati, Marco~G Genoni, Xian-Min Jin, W~Steven
  Kolthammer, MS~Kim, Animesh Datta, Marco Barbieri, and Ian~A Walmsley.
\newblock Joint estimation of phase and phase diffusion for quantum metrology.
\newblock {\em Nature communications}, 5(1):3532, 2014.

\bibitem{crowley2014}
Philip~JD Crowley, Animesh Datta, Marco Barbieri, and Ian~A Walmsley.
\newblock Tradeoff in simultaneous quantum-limited phase and loss estimation in
  interferometry.
\newblock {\em Physical Review A}, 89(2):023845, 2014.

\bibitem{pezze2017}
Luca Pezz{\`e}, Mario~A Ciampini, Nicol{\`o} Spagnolo, Peter~C Humphreys,
  Animesh Datta, Ian~A Walmsley, Marco Barbieri, Fabio Sciarrino, and Augusto
  Smerzi.
\newblock Optimal measurements for simultaneous quantum estimation of multiple
  phases.
\newblock {\em Physical review letters}, 119(13):130504, 2017.

\bibitem{gessner2018}
Manuel Gessner, Luca Pezz{\`e}, and Augusto Smerzi.
\newblock Sensitivity bounds for multiparameter quantum metrology.
\newblock {\em Physical review letters}, 121(13):130503, 2018.

\bibitem{proctor2018}
Timothy~J Proctor, Paul~A Knott, and Jacob~A Dunningham.
\newblock Multiparameter estimation in networked quantum sensors.
\newblock {\em Physical review letters}, 120(8):080501, 2018.

\bibitem{gefen2019}
Tuvia Gefen, Amit Rotem, and Alex Retzker.
\newblock Overcoming resolution limits with quantum sensing.
\newblock {\em Nature communications}, 10(1):4992, 2019.

\bibitem{chen2019}
Hongzhen Chen and Haidong Yuan.
\newblock Optimal joint estimation of multiple rabi frequencies.
\newblock {\em Physical Review A}, 99(3):032122, 2019.

\bibitem{PhysRevLett.134.030802}
Majid Hassani, Santiago Scheiner, Matteo G.~A. Paris, and Damian Markham.
\newblock Privacy in networks of quantum sensors.
\newblock {\em Phys. Rev. Lett.}, 134:030802, Jan 2025.

\bibitem{belliardo2024}
Federico Belliardo, Valeria Cimini, Emanuele Polino, Francesco Hoch, Bruno
  Piccirillo, Nicol{\`o} Spagnolo, Vittorio Giovannetti, and Fabio Sciarrino.
\newblock Optimizing quantum-enhanced bayesian multiparameter estimation of
  phase and noise in practical sensors.
\newblock {\em Physical Review Research}, 6(2):023201, 2024.

\bibitem{Humphreys2013}
Peter~C. Humphreys, Marco Barbieri, Animesh Datta, and Ian~A. Walmsley.
\newblock Quantum enhanced multiple phase estimation.
\newblock {\em Phys. Rev. Lett.}, 111:070403, Aug 2013.

\bibitem{yue2014}
Jie-Dong Yue, Yu-Ran Zhang, and Heng Fan.
\newblock Quantum-enhanced metrology for multiple phase estimation with noise.
\newblock {\em Scientific reports}, 4(1):5933, 2014.

\bibitem{liu2016}
Jing Liu, Xiao-Ming Lu, Zhe Sun, and Xiaoguang Wang.
\newblock Quantum multiparameter metrology with generalized entangled coherent
  state.
\newblock {\em Journal of Physics A: Mathematical and Theoretical},
  49(11):115302, 2016.

\bibitem{gagatsos2016}
Christos~N Gagatsos, Dominic Branford, and Animesh Datta.
\newblock Gaussian systems for quantum-enhanced multiple phase estimation.
\newblock {\em Physical Review A}, 94(4):042342, 2016.

\bibitem{holevo2011}
Alexander~S Holevo.
\newblock {\em Probabilistic and statistical aspects of quantum theory},
  volume~1.
\newblock Springer Science \& Business Media, 2011.

\bibitem{helstrom1969}
Carl~W Helstrom.
\newblock Quantum detection and estimation theory.
\newblock {\em Journal of Statistical Physics}, 1:231--252, 1969.

\bibitem{caves1}
Carlton~M. Caves.
\newblock Quantum-mechanical radiation-pressure fluctuations in an
  interferometer.
\newblock {\em Phys. Rev. Lett.}, 45:75--79, Jul 1980.

\bibitem{caves2}
Carlton~M. Caves.
\newblock Quantum-mechanical noise in an interferometer.
\newblock {\em Phys. Rev. D}, 23:1693--1708, Apr 1981.

\bibitem{braunstein1994}
Samuel~L Braunstein and Carlton~M Caves.
\newblock Statistical distance and the geometry of quantum states.
\newblock {\em Physical Review Letters}, 72(22):3439, 1994.

\bibitem{paris2009quantum}
Matteo G.~A. Paris.
\newblock Quantum estimation for quantum technology.
\newblock {\em International Journal of Quantum Information},
  7(supp01):125--137, 2009.

\bibitem{brown2003statistical}
Kevin~S Brown and James~P Sethna.
\newblock Statistical mechanical approaches to models with many poorly known
  parameters.
\newblock {\em Physical review E}, 68(2):021904, 2003.

\bibitem{brown2004statistical}
Kevin~S Brown, Colin~C Hill, Guillermo~A Calero, Christopher~R Myers, Kelvin~H
  Lee, James~P Sethna, and Richard~A Cerione.
\newblock The statistical mechanics of complex signaling networks: nerve growth
  factor signaling.
\newblock {\em Physical biology}, 1(3):184, 2004.

\bibitem{PhysRevLett.97.150601}
Joshua~J. Waterfall, Fergal~P. Casey, Ryan~N. Gutenkunst, Kevin~S. Brown,
  Christopher~R. Myers, Piet~W. Brouwer, Veit Elser, and James~P. Sethna.
\newblock Sloppy-model universality class and the vandermonde matrix.
\newblock {\em Phys. Rev. Lett.}, 97:150601, Oct 2006.

\bibitem{machta2013parameter}
Benjamin~B Machta, Ricky Chachra, Mark~K Transtrum, and James~P Sethna.
\newblock Parameter space compression underlies emergent theories and
  predictive models.
\newblock {\em Science}, 342(6158):604--607, 2013.

\bibitem{PRXQuantum.2.020308}
Lukas~J. Fiderer, Tommaso Tufarelli, Samanta Piano, and Gerardo Adesso.
\newblock General expressions for the quantum fisher information matrix with
  applications to discrete quantum imaging.
\newblock {\em PRX Quantum}, 2:020308, Apr 2021.

\bibitem{Goldberg21}
Aaron~Z. Goldberg, Jos\'{e}~L. Romero, \'{A}ngel~S. Sanz, and Luis~L.
  S\'{a}nchez-Soto.
\newblock Taming singularities of the quantum fisher information.
\newblock {\em International Journal of Quantum Information}, 19(08):2140004,
  2021.

\bibitem{Yang23}
Yu~Yang, Federico Belliardo, Vittorio Giovannetti, and Fuli Li.
\newblock Untwining multiple parameters at the exclusive zero-coincidence
  points with quantum control.
\newblock {\em New Journal of Physics}, 24(12):123041, jan 2023.

\bibitem{Frigerio2024}
Massimo Frigerio and Matteo G.~A. Paris.
\newblock {Overcoming sloppiness for enhanced metrology in continuous-variable
  quantum statistical models}.
\newblock {\em arXiv preprint arXiv:2410.02989}, oct 2024.

\bibitem{PhysRevA.111.012414}
Jiaxuan Wang and Girish~S. Agarwal.
\newblock Exact quantum fisher matrix results for distributed phases using
  multiphoton polarization greenberger-horne-zeilinger states.
\newblock {\em Phys. Rev. A}, 111:012414, Jan 2025.

\bibitem{zhu2015}
Huangjun Zhu.
\newblock Information complementarity: A new paradigm for decoding quantum
  incompatibility.
\newblock {\em Scientific reports}, 5(1):14317, 2015.

\bibitem{heinosaari2016}
Teiko Heinosaari, Takayuki Miyadera, and M{\'a}rio Ziman.
\newblock An invitation to quantum incompatibility.
\newblock {\em Journal of Physics A: Mathematical and Theoretical},
  49(12):123001, 2016.

\bibitem{ragy2016}
Sammy Ragy, Marcin Jarzyna, and Rafa{\l} Demkowicz-Dobrza{\'n}ski.
\newblock Compatibility in multiparameter quantum metrology.
\newblock {\em Physical Review A}, 94(5):052108, 2016.

\bibitem{candeloro2024dimension}
Alessandro Candeloro, Zahra Pazhotan, and Matteo G.~A. Paris.
\newblock Dimension matters: precision and incompatibility in multi-parameter
  quantum estimation models.
\newblock {\em Quantum Science and Technology}, 9(4):045045, 2024.

\bibitem{adani2024}
Marco Adani, Simone Cavazzoni, Berihu Teklu, Paolo Bordone, and Matteo G.~A.
  Paris.
\newblock Critical metrology of minimally accessible anisotropic spin chains.
\newblock {\em Scientific Reports}, 14(1):19933, 2024.

\bibitem{cavazzoni2024}
Simone Cavazzoni, Marco Adani, Paolo Bordone, and Matteo G.~A. Paris.
\newblock Characterization of partially accessible anisotropic spin chains in
  the presence of anti-symmetric exchange.
\newblock {\em New Journal of Physics}, 26(5):053024, 2024.

\bibitem{cramer1999}
Harald Cram{\'e}r.
\newblock {\em Mathematical methods of statistics}, volume~9.
\newblock Princeton university press, 1999.

\bibitem{kay1993}
Steven~M Kay.
\newblock Statistical signal processing: estimation theory.
\newblock {\em Prentice Hall}, 1:Chapter--3, 1993.

\bibitem{helstrom1967}
Carl~W Helstrom.
\newblock Minimum mean-squared error of estimates in quantum statistics.
\newblock {\em Physics letters A}, 25(2):101--102, 1967.

\bibitem{Yuen1973}
Horace Yuen and Melvin Lax.
\newblock Multiple-parameter quantum estimation and measurement of
  nonselfadjoint observables.
\newblock {\em IEEE Transactions on Information Theory}, 19(6):740--750, 1973.

\bibitem{FujiwaraA1994}
A.~Fujiwara.
\newblock Multi-parameter on the right pure state estimation based on the right
  logarithmic derivative.
\newblock Technical Report METR94-08, The University of Tokyo, Department of
  Mathematical Engineering and Information Physics, 1994.
\newblock Available at
  \url{http://www.keisu.t.u-tokyo.ac.jp/research/techrep/data/1994/METR94-09.pdf}.

\bibitem{Nagaoka2005}
Hiroshi Nagaoka.
\newblock A new approach to cram{\'e}r-rao bounds for quantum state estimation.
\newblock In {\em Asymptotic Theory of Quantum Statistical Inference: Selected
  Papers}, pages 100--112. WORLD SCIENTIFIC, 2005.

\bibitem{hayashi2008}
Masahito Hayashi and Keiji Matsumoto.
\newblock Asymptotic performance of optimal state estimation in qubit system.
\newblock {\em Journal of Mathematical Physics}, 49(10), 2008.

\bibitem{kahn2009}
Jonas Kahn and M{\u{a}}d{\u{a}}lin Gu{\c{t}}{\u{a}}.
\newblock Local asymptotic normality for finite dimensional quantum systems.
\newblock {\em Communications in Mathematical Physics}, 289(2):597--652, 2009.

\bibitem{Yamagata2013}
Koichi Yamagata, Akio Fujiwara, and Richard~D. Gill.
\newblock Quantum local asymptotic normality based on a new quantum likelihood
  ratio.
\newblock {\em The Annals of Statistics}, 41(4):2197 -- 2217, 2013.

\bibitem{Suzuki2016}
Jun Suzuki.
\newblock Explicit formula for the holevo bound for two-parameter qubit-state
  estimation problem.
\newblock {\em Journal of Mathematical Physics}, 57(4), 2016.

\bibitem{PhysRevA.91.042104}
S.~Alipour and A.~T. Rezakhani.
\newblock Extended convexity of quantum fisher information in quantum
  metrology.
\newblock {\em Phys. Rev. A}, 91:042104, Apr 2015.

\end{thebibliography}
\end{document}